\documentstyle[12pt]{article}
\input psfig
\long\def\comment#1{}
\begin{document}
\title{Uncertainty In  Quantum Computation}
\author{Subhash Kak\\
Louisiana State University\\
Baton Rouge, LA 70803, USA}
\maketitle

\begin{abstract}
We examine the effect of 
previous history on starting a computation on a
quantum computer.
Specifically, we assume that the quantum register
has some unknown state on it, and it is required that
this state be cleared and replaced by a specific 
superposition state without any phase uncertainty,
as needed by quantum algorithms.
We show that, in general, this task is computationally
impossible.

\end{abstract}

\section{Introduction}

Quantum computing algorithms (for example, \cite{Ek96,Gr97})
work in two ways: (i) they apply an appropriate
unitary transformation 
on a superposition state; (ii) they increase the amplitude of the state
that represents the solution to the problem so that,
upon interaction of a measuring device with the 
quantum register, the solution is available
with a high probability.
This conception proceeds without any uncertainty.

But there are uncertainties in all quantum description,
which is why quantum information is not subject to the
same rules as classical information\cite{Ka98}. 
When a measurement is made on one qubit (which may be imagined
to be a photon with an unknown angle of polarization)
that has not been
examined before, it is as likely
to be a 0 as a 1; when measured again from a
different orientation, its uncertainty remains. If
$\Delta I$ represents the change in information and 
$\Delta t$ represents the measurement step (counted in
integers), one may speak
of the following uncertainty constraint for information:

\[ \Delta I + \Delta t \geq 1 \]
This says that any unknown qubit will yield one bit of information,
and when checked again, the result may be the same, but at the
expense of the additional interaction. If observed frequently enough,
the state will become frozen (the quantum Zeno effect).

There is also uncertainty
related to the preparation of the initial state.
In earlier papers\cite{Ka99,Ka00,Ka01a},
it was argued that the quantum computing model leaves
out several elements of physical constraints and 
the problem of initial state preparation, making
the paradigm computationally unrealistic\cite{Ka01b}.
In this analysis, it was assumed that somehow the quantum register
will be constituted as qubits become available, one by one.

In the present paper,
we consider the problem from a slightly different perspective,
where the quantum register is supposed to have already been used
for the solution of some problem. The task now is to prepare this
register to start a new computation. The standard quantum algorithms
require that the register be in a definite superposition state with, at worst, 
an unknown global phase.

Since the quantum register -- howsoever it is implemented -- is
a single system, the question arises: Can one determine the
state of this system, so that the appropriate unitary 
transformation can be carried out on it to take it to
the desired initial state of the next computation?
According to quantum mechanics, it is 
impossible to determine the unknown quantum wavefunction of
a single system\cite{Al01},
so we must first interact with the
register to reduce its wavefunction to some convenient
eigenstate.
But steering this eigenstate to the desired starting point
of the computation is much more difficult than
may be imagined.
The precision of the macroscopic measurement device is
limited because its own state can never be completely
known and its interaction with the
quantum register represents a 
many-to-one mapping (many register states mapped to the
same measuring device state\cite{Ze70}).
We show that, in general, one {\it cannot} steer the
eigenstate to a specific superposition state without any unknown
phases between the components.

\section{The initial superposition state}

Let there be a total number of $n$ qubits.
Without going into the question of 
quantum statistical constraints\cite{Ka01a},
the total number of component (eigenstates) states of interest
from the point of view of computation is $N=2^n$.

When $n=2$, for example, the number of
component states is four, being
$|00\rangle, ~|01\rangle,~ |10\rangle,~ |11\rangle$.
These states may be represented by
$(1,0,0,0)$, $(0,1,0,0)$, $(0,0,1,0)$, and $(0,0,0,1)$,
and they 
may be
combined in an infinite variety of
combinations by the use of complex weights
to create superposition state functions.

The initial  state, $|R_i \rangle$,
is normally taken to be $(1, 0, 0...0)$, whereas
the final superposition state is taken to be
$(1, 1, 1...1)$.
However, no distinction is made between

\begin{equation}
 |R_i \rangle = (1, 1, 1, ..., 1)
\end{equation}
and

\begin{equation}
 |R_i \rangle = (1, e^{i \theta_1}, e^{i \theta_2}, ..., e^{i \theta_{N -1}})
\end{equation}
even though they are distinct states.

To eliminate states with nonzero $\theta_i$s, one confronts the
problem of estimating an unknown wavefunction, a problem that is insoluble.
What one can do is to test the wavefunction 
with an appropriate measuring device for its {\it a priori}
values.

For example, consider the states $(1, 1)$ and $(1, i)$.
If viewed as photons, they each represent polarization at
45$^o$. If we perform the following unitary transformation:

\begin{equation}
			 \left[ \begin{array}{cc}
                                  {\sqrt 3} /2  & -i/2 \\
                                  i/2 & -{\sqrt 3}/2 \\
                               \end{array} \right] 
\end{equation}
we are led to new superposition states with
equal probabilities for $ |0\rangle$ and $ | 1\rangle$ if the
starting state is $(1, 1)$; but probabilities of
$93 \%$ and $ 7 \%$ 
for $ |0\rangle$ and $ | 1\rangle$ if the
starting state is $(1, i)$.

Clearly, we must insist on a distinction between
various states, such as (1) and (2), if they have different phases amongst their
components.
But, there is no way of ensuring that a given superposition does not
have relative phases, as in (2), unless one does additional tests.

If one knew {\it a priori} what the relative phase was, then
one could remove it, as in the example where the
qubit is $(1, e^{i \theta})$. Such a qubit can be aligned back if the unitary 
transformation

\begin{equation}
			 \left[ \begin{array}{cc}
                                  1 /{\sqrt 2}  & e^{-i \theta}/{\sqrt 2} \\
                                  e^{i \theta}/{\sqrt 2} & -1/{\sqrt 2} \\
                               \end{array} \right] 
\end{equation}
is employed.
But since there is no way of knowing this unknown
$\theta$, such an operation cannot be performed.
It is a chicken-and-egg problem: If one knew what the
phases were, one could get rid of them, but there is no
way of {\it a priori} knowing these phases.

\section{Number of unitary operations}

The state of the  macroscopic measuring device, $M$, 
cannot be completely known\cite{Ze70}.
Upon measurement of (2) by it, 
the state on the register will be some random

 \[ ( 0, 0, ..., 1, ..., 0) \]
where the randomness is with regard to the location of the
1 in the vector.
In other words,
the initial measurement to ``erase'' the old
information on the quantum register
reduces the wavefunction to 
one of the $2^n$ component states.
One would need an appropriate
transformation out of a total of $2^n$ unitary
transformations to steer this component 
state into the specific starting state
of the quantum computer.

The cost of achieving the appropriate transformation
increases exponentially with respect to the number
of qubits.

If the erasing apparatus is not in {\it perfect} alignment with the
apparatus used to implement the unitary transformation
and the final measurement,
then we have a further complication.
The
reduced state will now be a superposition of the
components of the quantum register, with $N-1$
unknown phases. 
To steer this state to the starting superposition state
would be {\it impossible}, because the number of cases would
be infinite.

The measuring apparatus, $M$, and the quantum
register, $R$, have a joint state function that is
a product of the individual states. When
the measurement produces some
eigenvalue of $M$, the register wavefunction
is correspondingly
reduced. But the state of the macroscopic apparatus
can only be determined incompletely, therefore there will
be a residual uncertainty with regard to the knowledge
obtained about the quantum register.

\section{Conclusions}

We have shown that if the measurement apparatus is perfectly
aligned with the quantum register, the task of erasing the
old information requires the use of one of $2^n$ unitary
transformations.
When the alignment is not perfect, the task is impossible.
Since the measuring apparatus and the register cannot,
in general, be aligned {\it perfectly} because of
limitations of precision and noise, the problem of 
initializing the register is impossible to solve.

The information obtained about the register wavefunction 
is indirect, through observations on the macroscopic measuring
apparatus $M$. But since information about it must
remain incomplete, there exists a corresponding incompleteness
in our knowledge of the register wavefunction.

Alter and Yamamoto assert\cite{Al01}: ``The information that
can be obtained about the quantum wavefunction of a single
system in a series of measurements cannot account for the
physical reality (i.e., ontological meaning) of the
wavefunction, and that the quantum wavefunction is limited
to having a statistical (i.e., epistemological) meaning only.''
This is another way of looking at the difficulty of erasing past
information from a quantum register, emphasizing the fact that
the wavefunction provides meaning when used for an ensemble of
systems.

The precision with which one can know the wavefunction depends
on the maximum information that can flow from the system into
the measuring device.
As the device can get into one of $2^n$ states, the total
information that can be known about the wavefunction is
$log_2 2^n = n$ bits.
The total uncertainty associated with the wavefunction, given
that there exist $(N-1)$ relative phases of arbitrary value,
is without bound.
Therefore, it is impossible to completely know the register
wavefunction.
This is in accord with the observation of Einstein,
Tolman and Podolsky\cite{Ei31} that ``the principle of
the quantum mechanics must involve an uncertainty in the
description of the past events which is analogous to the
uncertainty in the prediction of future events.''

\end{document}